\documentclass[%
 reprint,
 amsmath,amssymb,
 prd,
]{revtex4-2}

\usepackage{graphicx}% Include figure files
\usepackage{makecell}
\usepackage{dcolumn}% Align table columns on decimal point
\usepackage{bm}% bold math
\usepackage[export]{adjustbox} % Para ajustar la posición de la figura
\usepackage{array}
\usepackage{amsmath}
\usepackage{amsfonts} 
\usepackage{float} 
\usepackage[caption=false]{subfig}
\usepackage[usenames,dvipsnames,svgnames,table]{xcolor}
\usepackage{hyperref}% add hypertext capabilities
\definecolor{verdeoscuro}{rgb}{0, 0.5, 0}

\begin{document}

\preprint{APS/123-QED}

\title{Enhancement of dark‑photon haloscope sensitivity with degenerate modes: \\ toward axion‑level form factor and polarization determination}
% Optimal enhancement of dark photon detection \\ using haloscopes with degenerate resonant modes
% Force line breaks with \\

\author{Jose R. Navarro-Madrid$^{a}$}
\thanks{Corresponding author: joser.navarro@upct.es} 
\author{José Reina-Valero$^{b}$} 
\author{Alejandro D\'iaz-Morcillo$^{a}$} 
\author{Benito Gimeno$^{b}$}

\affiliation{$^a$Departamento de Tecnolog\'ias de la Informaci\'on y las Comunicaciones,
Universidad Polit\'ecnica de Cartagena,
Plaza del Hospital 1, 30202 Cartagena, Spain. \\ 
$^b$Instituto de Física Corpuscular (IFIC), CSIC-University of Valencia, Calle Catedr\'atico Jos\'e Beltr\'an Martínez, 2, 46980 Paterna (Valencia), Spain.}

% E-mail addresses: only for the corresponding author

\date{\today}% It is always \today, today,

\begin{abstract}
The dark photon has been postulated as a potential constituent of dark matter, exhibiting notable similarities to the axion. The primary distinction between the two particles lies in the nature of their respective fields: the dark photon field is a vector field with a polarization direction that remains undetermined. This work explores the prospect of utilizing three degenerate modes for scanning the three dimensions of space in order to mitigate the low form factor expected in the detection of the dark photon due to their unknown polarization. The employment of an haloscope with three orthogonal and degenerate modes in conjunction with the coherent sum of signals is demonstrated in this work in order to enhance the dark photon form factor up to the axion form factor, and to determine the direction of the dark photon polarization vector. We show in this manuscript that the maximum form factor is achieved in cavities of cubic, spherical, and cylindrical geometries, considering the introduction of tuning elements. To achieve this adequately, some conditions reviewed in this article must be fulfilled in the resonant cavity, leading to uncertainties in the final measurement. Finally, this technique can allow the simultaneous search for dark matter axions and dark photons, and to the knowledge of the authors, the method shown in this work is the most effective one for detecting dark photon with microwave resonant cavities.
\end{abstract}

\maketitle

%\tableofcontents

%%%%%%%%%%%%%%%%%%%%%%%%%%%%%%%%%%%%%%%%%%%%%%%%%%%%%%%%%%%%
\section{Introduction} \label{sec:intro}
%%%%%%%%%%%%%%%%%%%%%%%%%%%%%%%%%%%%%%%%%%%%%%%%%%%%%%%%%%%%

The nature of Dark Matter (DM) can be explained with a particle model by several candidates. Probably, nowadays the most popular one would be the axion, originated by the spontaneous symmetry breaking of an additional U(1) symmetry of the Standard Model (SM), with the aim of giving an explanation to the suspicious cancellation of the $\overline{\theta}$ angle in the Charge-Parity (CP) violating term of Quantum ChromoDynamics (QCD) \cite{Peccei_Quinn, Weinberg, Wilczek}. Another candidate that has raised interest too is the Dark Photon (DP), originated as a very simple modification to the SM, very similar to the axion case, by just adding an additional U(1) symmetry that, when spontaneously broken, arises a massive vector particle, in this case the DP \cite{Dark_Photon_book, DP_handbook}. However, since this particle is not related to any additional problem, is the axion where the scientific community tends to put the focus into \cite{ADMX2018.24, CAPP2021, RADES_SM18}. \\

Commonly, a portal through the electromagnetic (EM) interaction is searched to access to this hidden sector of matter. In the DP case, this portal comes as a kinetic mixing with the ordinary photon, as it can be seen in Eq. \ref{eq:lagrangian} (expressed in natural units)\footnote{All equations in this article are expressed in SI units unless otherwise specified.}

\begin{equation}
\begin{aligned}
    \mathcal{L} \supset -\frac{1}{4} F_{\mu \nu } F^{\mu \nu} - \frac{1}{4} X_{\mu \nu} X^{\mu \nu} + \frac{\chi}{2} F_{\mu \nu} X^{\mu \nu} \\
     +\frac{m_X^2 }{2}X_\mu X^\mu + e J_{EM}^\mu A_\mu,
    \label{eq:lagrangian}
\end{aligned}
\end{equation}
where $F^{\mu\nu}$ is the Faraday tensor, $X^{\mu\nu}$ is the strength tensor of the DP, $A^{\mu}$ and $X^{\mu}$ are the photon and DP fields, respectively; $\chi$ is the DP mixing constant, $m_{X}$ the DP mass, $e$ the electron charge, $J^{\mu}_{EM}$ the equivalent EM electric current density, and the Einstein indexes summation criterion has been used in this expression. In this way, the same setups for detecting the QCD axion (since as it is well-known, it might couple to the ordinary photon \cite{Sikivie, Sikivie_2}) can be employed for DP detection, in which our focus is put into. However, while in the axion case an external intense static magnetic field is needed due to the structure of the interaction term, this is not mandatory for the DP. Due to this simultaneous search of both axions and DPs when a new bound is established by an axion experiment to the coupling constant $g_{a\gamma\gamma}$, this bound can be translated as well to the mixing constant of the DP, $\chi$ \cite{AxionLimits}. \\

The expression of the spatial components of the DP field is given by \cite{Dark_Photon_book}
\begin{equation}
    \vec{X} = \hat{n} \frac{\sqrt{2\rho_0}}{m_{X}},
\end{equation}
where $\rho_0$ is the DM density on the Earth and $\hat{n}$ is the intrinsic polarization vector as happens with the ordinary photon, which satisfies $||\hat{n}||=1$. The latter feature of the DP introduces some peculiarities in its detection process in comparison with the DM axion. More specifically, since this polarization comes from cosmological constraints of the primordial universe, it remains a prior unknown. Due to this, different assumptions can be followed for the theoretical calculation of the DP form factor, implying in any case a significative decrease of this magnitude in comparison with DM axions detection. In this way, regarding previous work where multimode detection in an haloscope is used \cite{Pablo_HFGW, JR_HFGW}, a frequency degenerate multimode haloscope is proposed for solving this issue with the aim of maximizing the DP form factor. \\

This article is organized as follows: first, a description of the experimental parameters in DP detection is performed, focusing on the form factor difference with respect to the axion case; second, the possibility of coherently summing signals of three degenerate modes to obtain the maximum form factor is demonstrated by using a full-wave modal technique commonly used in the context of advanced microwave engineering. Then, the development of a formulation that takes into account possible errors in the phase sum due to irregularities in the cavities fabrication is performed. Next, a description of three different canonical geometries for resonant cavities is done, including the analysis of the related resonant modes. The form factor is calculated for the analyzed cases considering all possible directions of the DP polarization vector. Finally, the improvement factor in both sensitivity and integration time is obtained for each cavity.

%%%%%%%%%%%%%%%%%%%%%%%%%%%%%%%%%%%%%%%%%%%%%%%%%%%%%%%%%%%%
\section{Dark photon form factor}\label{sec:formfactor}
%%%%%%%%%%%%%%%%%%%%%%%%%%%%%%%%%%%%%%%%%%%%%%%%%%%%%%%%%%%%
%One of the major differences between the axion and the DP is that while the axion consists of a scalar field, the DP field has an intrinsic polarization $\hat{n}$, and the direction to which it points is unknown. 

An important experimental parameter is the detected power $P_d$ generated by the DP within the haloscope. In a haloscope based on a resonant cavity, the conventional expression of this power at the resonant frequency is given by \cite{WISPY}
\begin{equation} \label{eq:Pd}
    P_{d} = \kappa\chi^2 m_{X} \rho_0 Q_L V C_\mathrm{DP},
\end{equation}
where $\kappa$ is the coupling factor of the port, $Q_L$ is the loaded quality factor, $V$ is the volume of the cavity, and $C_\mathrm{DP}$ is the DP form factor; we have assumed that $Q_L<<Q_\mathrm{DP}$, being $Q_\mathrm{DP}$ the DP quality factor. The form factor quantifies the coupling between the DP and the modal resonant electric field ($\vec{E}$) of the mode excited by the DP in the cavity as follows
\begin{equation} \label{eq:formfactordef}
    C_\mathrm{DP}=\frac{|\int_V \vec{E}\cdot\hat{n}\:dV|^2}{V \int_V\|\vec{E}\|^2\:dV}.
\end{equation}

Similarly to the axion case \cite{RADESuniverse}, the sensitivity that the experiment can reach is given by 
\begin{equation} 
    \chi = \left(\frac{\frac{S}{N} k_B T_{sys}}{\kappa \rho_0 C_{\mathrm{DP} }V Q_L} \right)^\frac{1}{2} \left( \frac{1}{ m_X Q_{\mathrm{DP}}\Delta t} \right)^\frac{1}{4}.
    \label{eq:chi}
\end{equation}
where $\frac{S}{N}$ is the signal-to-noise ratio, $k_B$ is the Boltzmann constant, $T_{sys}$ is the system noise temperature and $\Delta t$ is the integration time. Such equation confirms that for the case of DPs, in order to obtain better sensitivities, the parameters that can be optimized in the design of the cavity are the unloaded quality factor $Q_L$, $V$ and $C_{\mathrm{DP}}$. Thus, it can be confirmed that the form factor optimization plays an important role for improving the detectability of DPs. \\

Equation \ref{eq:formfactordef} is analogous to the expression for the axion form factor (see for instance \cite{RADESuniverse}), but in this case the direction of the external magnetostatic field is replaced by the direction of the DP polarization vector $\hat{n}$. Therefore, the DP form factor can be expressed as \cite{WISPY}
\begin{equation}
    C_\mathrm{DP} = C_{a}\cos^2({\theta}),
\end{equation}
where $C_a$ is the axion form factor, and $\theta$ is the angle between the external static magnetic field applied in the haloscope (for the axion detection) and the direction of $\hat{n}$. It is illuminating to reemphasize that for the DP detection, no external magnetostaic field is needed. Hence, the lack of knowledge of $\hat{n}$ arises two different approaches for facing this problem from both experimental and theoretical point of views \cite{WISPY}. The first one is that the direction of the DP field is not affected by the DM distribution and the DPs point to the same direction \footnote{In this case, the DP polarization remains constant with respect to the galactic reference frame, but one should take into account that the laboratory reference frame is movable, thus varying the DP polarization as well, depending on the periodic movement of the reference frame.} (at least in a sufficiently large region of space), and the second is that the DPs behave like an ideal gas, thus having a variable polarization and therefore the particles point towards random directions, being this one a common approach followed in previous publications \cite{Bajjali_2023,PhysRevLett.130.181001, Nguyen_2019, PhysRevD.110.043022, PhysRevD.109.012008}. In any case, these two approaches lead to a reduction of the form factor with respect to the maximum value of the axion form factor achievable for a proper chosen EM resonant mode. \\

Assuming that $\hat{n}$ is static, a conservative approximation on the value of $\cos^2({\theta})$ is to estimate that the real value is greater than the expected one with 95\% probability, and therefore, it adopts a value of $\cos^2({\theta}) = 0.0025$. In the situation of a random direction and variable $\hat{n}$, the space average $\langle \rangle$ of possible directions gives $\langle \cos^2({\theta})\rangle = 1/3$ \cite{WISPY}. In both scenarios, a significant decrease in the form factor of the experiment is observed due to these values of $C_{\mathrm{DP}}$. This is a clear disadvantage for the simultaneous detection of DPs in axion detection experiments using only the corresponding single axion resonant mode. \\

The proposed solution in this work is to improve the form factor of the DP, regardless of the direction of $\hat{n}$, employing a haloscope with three degenerate modes whose electric field components cover all three possible Cartesian directions. Even though the DP coupling to each one of the three orthogonal modes is different, the coherent sum of the signal from the three waveguide ports may result in the maximum form factor possible, which is $C_a$, independently of the direction of $\hat{n}$, as we will demonstrate in this manuscript. This kind of enhancement of the DP detectability was also proposed in \cite{DP_handbook}. It is clear that, in order to take advantage of this method, the degenerate modes must be at exactly the same frequency, which will ensure that our haloscope has directional sensitivity in all possible directions of $\hat{n}$. In order to achieve this goal, the resonant cavities must incorporate high-precision tuning systems capable of slightly modify the resonant frequency of three degenerate modes independently for obtaining a good mechanical tuning. \\

Another benefit of this technique is the capability to determine the direction of $\hat{n}$ based on the power received at each port, and for this purpose, it is essential to measure the signals from the three orthogonal ports independently, as it will be discussed later.

%%%%%%%%%%%%%%%%%%%%%%%%%%%%%%%%%%%%%%%%%%%%%%%%%%%%%%%%%%%%%%%%%%%%%%%%%%%%%%%%%%%%%%%%%%%%%%%%%%%%
\section{Theoretical Formulation} \label{sec:birme}
%%%%%%%%%%%%%%%%%%%%%%%%%%%%%%%%%%%%%%%%%%%%%%%%%%%%%%%%%%%%%%%%%%%%%%%%%%%%%%%%%%%%%%%%%%%%%%%%%%%%
\subsection{Application of the BI-RME 3D method}
In this section we demonstrate that the signals extracted from the waveguide ports of the cavities can be coherently summed for any given direction of the polarization vector $\hat{n}$, achieving the maximum form factor possible for the case of the DP, which is $C_a$. For such purpose, the BI-RME 3D (Boundary Integral-Resonant Mode Expansion) method has been properly used. This technique was developed during the eighties and nineties of the last century at the Università degli Studio di Pavia (Italy), and it can be applied for obtaining the modal EM field distributions inside an arbitrarily-shaped cavity with a given number of ports connected to it, reporting information about both magnitude and phase of the involved input and output signals in a large bandwidth frequency band (not only in the resonant frequencies as provided by the conventional detected power formula in Eq. (\ref{eq:Pd})). The BI-RME 3D method has been explained in several publications \cite{advanced_modal_analysis, bi-rme, bi-rme_Pablo}, and here just the useful concepts for the analysis of the DP detection with haloscopes are directly applied. BI-RME 3D is formulated in SI units.\\

The BI-RME 3D method states that, given a cavity with an arbitrary number of ports $P$ connected to it, the equivalent currents measured in each waveguide port of the cavity (generated by a given internal source, which in this case is the DP signal) are expressed as
\begin{equation}
\begin{aligned}
     I_{l}^{(\mu)} = &\sum_{\nu = 1}^{P}\sum_{n=1}^{+\infty}Y_{ln}^{(\mu, \nu)}V_{n}^{(\nu)} \\
     &+\sum_{m=1}^{+\infty}F_{ml}^{(\mu)}\frac{\kappa_{m}}{\kappa_{m}^2 - k^2}\int_{V}\vec{E}_{m}\left(\vec{r}\,^\prime\right)\cdot \vec{J}_{\mathrm{DP}}\left(\vec{r}\,^\prime\right)\ dV^\prime,
     \label{eq:birme_current}
\end{aligned}
\end{equation}
where the indexes $\mu$ and $\nu$ make reference to a given port coupled to the cavity, and the indexes $l$ and $n$ make reference to the mode considered in such port; $Y_{ln}^{(\mu, \nu)}$ is the multimode generalized admittance matrix; $V_{n}^{\left(\nu\right)}$ is the equivalent voltage detected in the port $\nu$ of the port mode $n$; the index $m$ is related with the $m$-th resonant mode of the cavity; $\kappa_m$ are the perturbed eigenvalues of the lossy cavity, defined by
\begin{equation}
	\kappa_m  \, \approx  \,   k_m \, \left(1 - \frac{1}{2 \, Q_m}\right) \, + \, \mathrm{i} \, \frac{k_m}{2 \, Q_m}
	\nonumber     
\end{equation}
where $\mathrm{i}$ is the imaginary unit ($\mathrm{i}\equiv \sqrt{-1}$), $k_{m}$ is the wave-number of the $m$-th resonant mode of the lossless closed cavity, $k=\omega/c$ is the wave-number ($c$ is the speed of light in vacuum and $\omega=2\pi f$ is the angular frequency), and $Q_m$ the unloaded quality factor of the $m$-th resonant mode of the cavity. Also, the term $F_{ml}^{(\mu)}$ in Eq. \ref{eq:birme_current} has the following structure
\begin{equation}
    F_{ml}^{(\mu)} \equiv \int_{S(\mu)}\vec{H}_{m}\left(\vec{r}\right)\cdot \vec{h}_{l}^{(\mu)}\left(\vec{r}\right)\ dS,
\end{equation}
$\vec{H}_{m}$ being the normalized magnetic field of the $m$-th resonant mode of the cavity, and $\vec{h}_{l}^{\left(\mu\right)}$ is the normalized magnetic field of the $l$-th mode of the port $\mu$. Hence, this is the coupling integral between both magnetic fields in the contact surface between the port and the cavity. We want to remark that this surface integral accounts for the coupling between the cavity and the external waveguide ports. Finally, we can find the coupling integral between the normalized electric field of the $m$-th cavity normalized resonant mode $\vec{E}_{m}$ and the equivalent electric current density that excites the cavity $\vec{J}_\mathrm{DP}$ over its whole volume. One thing that has to be noted is that the electric and magnetic eigenvectors of the cavity are orthonormalized as follows
\begin{equation}
    \int_{V}\vec{E_i}\cdot\vec{E}_{j}\, dV = \delta_{ij}\ ; \ \ \ \int_{V}\vec{H_i}\cdot\vec{H}_{j}\, dV = \delta_{ij},\
\end{equation}
where $\delta_{ij}$ is the Kronecker delta symbol. The electric and magnetic eigenvectors are related by $\nabla \times \vec{E}_ m = k_m \vec{H}_ m$ and $\nabla \times \vec{H}_ m = k_m \,\vec{E}_ m$. Particularizing for the studied cavities, where $P=3$ ports are being considered $\mu, \nu \in \{1,2,3\}$, and that only the fundamental mode in the three waveguide ports is accounted ($n=l=1$), BI-RME 3D states that the DP acts as a time-harmonic current source in each port as follows
\begin{equation}
    I_{\mathrm{DP}_\mu} \, \equiv \, \sum_{m=1}^{M}F_{m1}^{(\mu)}\frac{\kappa_{m}}{\kappa_{m}^2 - k^2}\int_{V}\vec{E}_{m}\left(\vec{r}\,^\prime\right)\cdot \vec{J}_{\mathrm{DP}}\left(\vec{r}\,^\prime\right)\ dV^\prime
\end{equation}
where the infinite summation of the second term of Eq. \ref{eq:birme_current} has been reduced to the finite number of resonant modes $M$ requested in the simulated frequency band. \\

\begin{figure}[t]
    \centering
    \includegraphics[scale = 0.38]{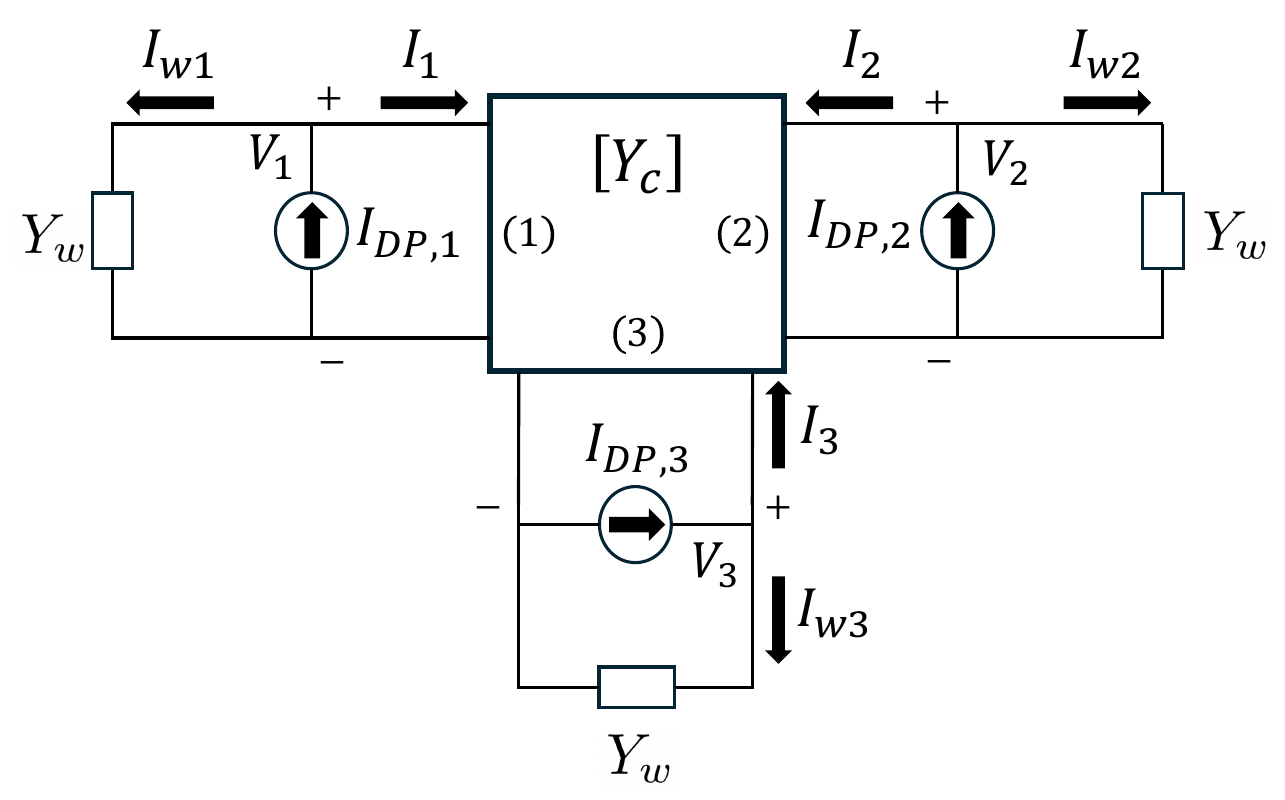}
    \caption{Equivalent network representing the DP decay inside the cavity.}
    \label{fig:RF_network}
\end{figure}

Next step is to rewrite Eq. (\ref{eq:birme_current}) in the following way:
\begin{equation}
      I_{\mathrm{DP}_\mu} =   I_{w_\mu} + I_\mu,
      \label{eq:Kirchhoff}
\end{equation}
with $I_{w_\mu} \equiv I_{1}^{\left(\mu\right)}$ and $I_\mu = \sum_{\nu = 1}^{3}Y_{11}^{(\mu, \nu)}V_{1}^{(\nu)}\equiv \sum_{\nu=1}^3 Y_{\mu\nu}V_{\nu}$, where, in order to alleviate the notation, the indexes relative to the waveguide modes have been eliminated. A simple interpretation of Eq. (\ref{eq:Kirchhoff}) with the well known first Kirchhoff law elucidates that the driven current source generated by the DP in each port is divided into two terms: one term is delivered to the ports $I_{w_{\mu}}$ for the detection of the signals, and the other currents $I_{\mu}$ enter into the cavity and are lost in form of ohmic losses (Joule effect), as it is shown in the single-mode network of Fig. \ref{fig:RF_network}. \\

Employing the basic formulation for the resolution of microwave networks based on the Kirchhoff laws, one can end up with the expression of the equivalent detected voltages in the ports $V_{\mu}$ in both magnitude and phase. Because of brevity, it is convenient to use matrix formulation. To this end, we define the current and voltage vectors, and the admittance matrices as follows
\begin{equation*}
    \bar{I} = \begin{pmatrix}
        I_{1}\\
        I_{2}\\
        I_{3}
    \end{pmatrix}; \ \ \ \bar{I}_{w} = \begin{pmatrix}
        I_{w_1}\\
        I_{w_2}\\
        I_{w_3}
    \end{pmatrix}; \ \ \ \overline{V} = \begin{pmatrix}
        V_{1}\\
        V_{2}\\
        V_{3}
    \end{pmatrix};
\end{equation*}
\begin{equation*}
    \bar{I}_{\mathrm{DP}} = \begin{pmatrix}
        I_{\mathrm{DP_1}}\\
        I_{\mathrm{DP_2}}\\
        I_{\mathrm{DP_3}}
    \end{pmatrix}; \ \ \ \overline{\overline{Y}}_c = \begin{pmatrix}
        Y_{11} & Y_{12} & Y_{13}\\
        Y_{21} & Y_{22} & Y_{23}\\
        Y_{31} & Y_{32} & Y_{33}
    \end{pmatrix};
\end{equation*}
\begin{equation}
    \overline{\overline{Y}}_w = \begin{pmatrix}
        Y_{w} & 0 & 0\\
        0 & Y_{w} & 0\\
        0 & 0 & Y_{w}
    \end{pmatrix}
\end{equation}
where $I_{w_\mu} = V_{\mu}\,Y_{w}$ is fulfilled, assuming that all ports are identical and present the same modal admittance $Y_w$. At this point, we are able to calculate the power extracted in each port, expressed as
\begin{equation}
    P_{w_\mu} = \frac{1}{2}\mathrm{Re}\left(V_\mu\,I_{w_\mu}^*\right)=\frac{1}{2}\left|V_\mu\right|^2 \mathrm{Re}(Y_{w}^*),
\end{equation}
where $*$ is complex conjugate, and hence the sum of powers of the three ports is given by
\begin{equation}
    P_{T} = \sum_{i=1}^3 P_{w_\mu} = \frac{1}{2}\mathrm{Re}\left(Y_{w}^{*}\right)\left[\left|V_1\right|^2 + \left|V_2\right|^2 + \left|V_3\right|^2\right].
    \label{eq:P_t}
\end{equation}

The DP current can be expressed as
\begin{equation}
\begin{aligned}
    \bar{I} &= \overline{\overline{Y}}_c\,\overline{V} = \bar{I}_{\mathrm{DP}} - \bar{I}_{w} = \bar{I}_{\mathrm{DP}} - \overline{\overline{Y}}_{w}\overline{V} \\
    \\
    &\Rightarrow \bar{I}_{\mathrm{DP}} = \left[\overline{\overline{Y}}_c + \overline{\overline{Y}}_w\right]\, \overline{V},
\end{aligned}
\end{equation}
ending up with the final expression for the voltages as a function of the excitation equivalent currents generated by the DP
\begin{equation} \label{eq:voltage}
    \overline{V} = \left[\overline{\overline{Y}}_{c} + \overline{\overline{Y}}_{w}\right]^{-1}\, \bar{I}_{\mathrm{DP}} = \bar{\bar{Z}}_{\alpha}\, \bar{I}_{DP}.
\end{equation}
where $\bar{\bar{Z}}_{\alpha} \equiv \left[\overline{\overline{Y}}_{c} + \overline{\overline{Y}}_{w}\right]^{-1}$ has been introduced to leverage the notation. Thus, it is observed that the phase of the detected voltage will only depend on the admittance matrix of the cavity and on the currents generated by the DP. In principle, if all ports are equal and the device is symmetrical and reciprocal, the admittance matrix does not introduce any relative phase between ports. \\

This let us with the $I_{\mathrm{DP_\mu}}$ being the only possible source of relative phase among the three ports. With the aim of developing a complete analysis, the equivalent electric current density generated by the DP has to be calculated. For doing so, we can employ the Lagrangian density in the interaction basis, neglecting terms of order $\mathcal{O}\left(\chi^2\right)$ (in natural units),
\begin{equation}
\begin{aligned}
    \mathcal{L} \supset &-\frac{1}{4} F_{\mu \nu } F^{\mu \nu} - \frac{1}{4} X_{\mu \nu} X^{\mu \nu}  + e J_{EM}^\mu A_\mu\\
    & + \frac{m_{X}^2}{2}\left(X^{\mu}X_{\mu} + 2\chi X_{\mu}A^{\mu}\right),
\end{aligned}
\end{equation}
where the Einstein indexes summation criterion has been used in this expression. Applying the Euler-Lagrange equations, one can obtain the inhomogeneous Maxwell equations modified by the presence of the DP
\begin{equation}
    \vec{\nabla}\cdot \vec{E} = m_{X}^2\chi X^{0},
\end{equation}
\begin{equation}
    \vec{\nabla}\times \vec{B} = \frac{\partial \vec{E}}{\partial t} - m_{X}^2\chi \vec{X},
\end{equation}
thus the equivalent charge and current densities of the DP are
\begin{equation}
    \rho_{\mathrm{DP}} = m_{X}^2\chi X^{0}; \ \ \ \vec{J}_{\mathrm{DP}} = - m_{X}^2\chi \vec{X} = -\omega_{X}\chi \sqrt{2 \rho_{0}}\,\hat{n}.
    \label{eq:J_equiv}
\end{equation}

This implies that the DP signals in each port have the same phase, and that when calculating $I_{\mathrm{DP_{\mu}}}$ with Eq. \ref{eq:birme_current} no relative phase is introduced among the coaxial ports, and therefore, we are able to sum the signals coherently.\\

Let us now develop the expression of the total power $P_{T}$ in order to demonstrate that the achievable form factor in our setup is the one of the axion. Combining \ref{eq:P_t} and \ref{eq:voltage}, we get
\begin{equation}
    P_T = \frac{1}{2}\mathrm{Re}(Y_{w}^*)\sum_{\mu,\nu=1}^3\left|Z_{\alpha_{\mu \nu}}\,I_{\mathrm{DP}_\nu}\right|^2.
    \label{eq:P_T_partial}
\end{equation}

For the moment, let us extract from the sum the factors related with $\bar{\bar{Z}}_{\alpha}$ as a common factor $\left|Z_{\alpha}^{D}\right|^2$. In order to do this, the admittance matrix of the cavity must fulfill some conditions, which will be explained in Section \ref{sec:error_estimation} below. Therefore, the total power can be expressed as
\begin{equation}
\begin{aligned}
       P_T = &\frac{1}{2} \mathrm{Re}(Y_w^*)\left|Z_{\alpha} ^D\right|^2\\
       &\times\sum_{\mu=1}^3\left|\sum_{m=1}^{M}F_{m1}^{(\mu)}\frac{\kappa_{m}}{\kappa_{m}^2 - k^2}\int_{V}\vec{E}_{m}\cdot\vec{J}_{\mathrm{DP}}\,dV\right|^2.
\end{aligned}
\end{equation}

The sum over the resonant modes of a completely symmetric cavity requests to consider the three degenerate modes ($M=3$). However, this summation can be truncated up to only one resonant mode, which in each port would be the resonant mode that couples to the given port, since the values of the other two integrals $F_{m1}^{(\mu)}$ are negligible in comparison with the main one. In addition, the integrals $F_{11}^{(\mu)}$ are equal in our case. Thus, the subindex $m$ can be dropped, only maintaining the index $\mu$, and the final expression of the total power, including the complete expression of $\vec{J}_{\mathrm{DP}}$ from Eq. (\ref{eq:J_equiv}) (transformed to SI units), is
\begin{equation}
\begin{aligned}
 P_T = &\mathrm{Re}(Y_{w}^*)\left|Z_{\alpha}^D\right|^2\\
 &\times\left|F_1\right|^2\left|\frac{\kappa_{1}}{\kappa_{1}^2 - k^2}\right|^2 \frac{1}{\mu_0}\omega_{X}^2\chi^2\rho_{0}\sum_{\mu=1}^3\left|\int_{V}\vec{E}_\mu\cdot\hat{n}\, dV\right|^2,
\label{eq:P_t_compensation}
\end{aligned}
\end{equation}
where $F_1 \equiv F_{11}^{(\mu)}$ and $\vec{E}_{\mu}$ is the normalized electric field of the $\mu$ resonant cavity mode.\\

From a practical point of view, three coaxial ports will be used in our proposal, which means that only the TEM fundamental coaxial mode will be considered. For this kind of antennas, the electric field of the three degenerate resonant modes points towards the three Cartesian axis of a reference frame located in the mass center of the cavity. If we would consider an axion experiment for each one of the modes separately, the DP form factors of each mode for a given direction of $\hat{n}$ could be expressed, in terms of the two conventional spherical angles elevation ($\theta$) and azimuth ($\varphi$), as
\begin{subequations}\label{eq:C_DP_components}
\begin{alignat}{2}
  C_{\mathrm{DP}_x} &= C_a\,\sin^2\theta\,\cos^2\varphi,\\ \quad
  C_{\mathrm{DP}_y} &= C_a\,\sin^2\theta\,\sin^2\varphi,\\
  C_{\mathrm{DP}_z} &= C_a\,\cos^2\theta.
\end{alignat}
\end{subequations}
and therefore $C_{\mathrm{DP}_x}+C_{\mathrm{DP}_y}+C_{\mathrm{DP}_z} = C_a$. As a consequence, this can be introduced into Eq. (\ref{eq:P_t_compensation}), thus obtaining
\begin{equation}
    P_T = \mathrm{Re}(Y_{w}^*)\left|Z_{\alpha}^D\right|^2\left|F_1\right|^2\left|\frac{\kappa_{1}}{\kappa_{1}^2 - k^2}\right|^2 \frac{1}{\mu_0}\omega_{X}^2\chi^2\rho_{0}VC_{a},
\end{equation}
having demonstrated that the form factor of the axion can be achieved in this setup by means of the employment of three degenerate and orthogonal resonant modes of a symmetric and reciprocal cavity resonator. \\

%%%%%%%%%%%%%%%%%%%%%%%%%%%%%%%%%%%%%%%%%%%%%%%%%%%%%%%%%%%%%%%%%
\subsection{Analysis of the error estimation in the coherent addition of the signals}\label{sec:error_estimation}
%%%%%%%%%%%%%%%%%%%%%%%%%%%%%%%%%%%%%%%%%%%%%%%%%%%%%%%%%%%%%%%%%%

In this subsection, an estimation of the error made when coherently summing the signals extracted from the three ports is developed. After Eq. (\ref{eq:P_T_partial}) the factors related with $\bar{\bar{Z}}_{\alpha}$ were extracted from the sum. This ideal summation case in which the form factors compensate between each other occurs only when the inverse of the admittance matrix fulfills two conditions:\\

$(i)$ The off‑diagonal elements are much smaller than the diagonal elements,\\

$(ii)$ Its diagonal elements are equal.\\

These conditions are fulfilled in a completely symmetrical and reciprocal cavity in terms of both geometry and waveguide ports. It is possible to deduce these conditions by examining expressions (\ref{eq:P_t}) and (\ref{eq:voltage}): at the end, the aim might be to find the quadratic sum of the elements of $\bar{I}_{\mathrm{DP}}$, since
\begin{equation}
    \left|I_{\mathrm{DP}_\mu}\right|^2 \propto \left|\int_{V}\vec{E}_{\mu}\cdot \hat{n}\ dV\right|^2.
\end{equation}
This case occurs when both mentioned conditions can be satisfied. However, the asymmetries imposed by the coaxial ports and the tuning systems (required to match resonant-mode frequencies), together with the differing quality factors of the modes, may prevent these conditions from being matched. \\

To analyze these situations, we will first address the error introduced by the fact that the off-diagonal terms of the admittance matrix are not as small as they should be, i.e, condition $(i)$. One can rewrite the voltages quadratic sum of Eq. (\ref{eq:P_t}) as follows
\begin{widetext}
\begin{equation}
\begin{aligned}
    \sum_{i=1}^{3}\left|V_i\right|^2&=\sum_{i=1}^3 \left|\sum_{j=1}^3 Z_{\alpha_{ij}}I_{\mathrm{DP}_j}\right|^2= \sum_{i=1}^3\sum_{j=1}^3\sum_{k=1}^{3} Z_{\alpha_{ij}}I_{\mathrm{DP}_j}Z_{\alpha_{ik}}^* I_{\mathrm{DP}_k}^{*} \left(\delta_{ij}\delta_{ik} + \left(1 - \delta_{ij}\delta_{ik}\right)\right)\\
 &=\sum_{i=1}^{3}\left|Z_{\alpha_{ii}}I_{\mathrm{DP}_i}\right|^2 + \sum_{i=1}^3\sum_{j=1}^3\sum_{k=1}^3 Z_{\alpha_{ij}}I_{\mathrm{DP}_j}Z_{\alpha_{ik}}^{*}I_{\mathrm{DP}_{k}}^{*}\left(1 - \delta_{ij}\delta_{ik}\right).
\end{aligned}
\end{equation}
\end{widetext}

In this way, the terms that for the moment do not introduce error (diagonal) have been separated from those that do introduce error (off-diagonal). Now, the sum relative to the terms within the diagonal can be extracted as a common factor, thus obtaining

\begin{widetext}
\begin{equation}
    \sum_{i=1}^3 \left|V_i\right|^2 = \sum_{i=1}^3 \left|Z_{\alpha_{ii}}I_{\mathrm{DP}_i}\right|^2 \left[1 + \frac{\sum_{l=1}^3\sum_{j=1}^3\sum_{k=1}^{3}Z_{\alpha_{lj}}Z_{\alpha_{lk}}^{*}I_{\mathrm{DP}_j}I_{\mathrm{DP}_k}^{*}\left(1 - \delta_{lj}\delta_{lk}\right)}{\sum_{l=1}^3\left|Z_{\alpha_{ll}}I_{\mathrm{DP}_l}\right|^2}\right] = \sum_{i=1}^3 \left|Z_{\alpha_{ii}}I_{\mathrm{DP}_i}\right|^2 \left(1 + \eta\right),
    \label{eq:error_fuera_diagonal}
\end{equation}
\end{widetext}
where $\eta$ is a term that quantifies the error due to consider that the off-diagonal terms are not negligible. The remaining task is to add the error resulting from unequal diagonal elements, i.e. condition $(ii)$. Considering an arbitrary element from the diagonal, $Z_{\alpha}^D$, and multiplying and dividing Eq. (\ref{eq:error_fuera_diagonal}) by $|Z_{\alpha}^D|^2$, it gets
\begin{equation}
    \sum_{i=1}^3\left|V_i\right|^2 =\left|Z_{\alpha}^D\right|^2\sum_{i=1}^3\frac{\left|Z_{\alpha_{ii}}\right|^2}{\left|Z_{\alpha}^D\right|^2}\left|I_{\mathrm{DP}_i}\right|^2\left(1+\eta\right). 
\end{equation}

Now we can rename the quotient $|Z_{\alpha_{ii}}|^2/|Z_{\alpha}^D|^2\equiv 1 + \zeta_i$. Thus, if the diagonal elements are equal to each other, $\zeta_i = 0,   \forall\,i$. Therefore,

\begin{equation}
   \sum_{i=1}^3\left|V_i\right|^2
=\left|Z_{\alpha}^D\right|^2\sum_{i=1}^3\left|I_{\mathrm{DP}_i}\right|^2\left(1 + \eta + \zeta_i\left(1 + \eta\right)\right).
\end{equation}

This expression can be separated into two parts:

\begin{widetext}
\begin{equation}
\begin{aligned}
\sum_{i=1}^3\left|V_i\right|^2 &= \sum_{i=1}^3\left|Z_{\alpha}^D\right|^2\left|I_{\mathrm{DP}_i}\right|^2 + \sum_{i = 1}^3 \left|Z_{\alpha}^D\right|^2\left|I_{\mathrm{DP}_i}\right|^2\left(\eta + \zeta_i\left(1 + \eta\right)\right)=  \sum_{i=1}^3\left|Z_{\alpha}^D\right|^2\left|I_{\mathrm{DP}_i}\right|^2 \left[1 + \frac{\sum_{j=1}^3\left|I_{\mathrm{DP}_j}\right|^2\left(\eta + \zeta_j\left(1 + \eta\right)\right)}{\sum_{j=1}^3 \left|I_{\mathrm{DP}_j}\right|^2}\right]\\
&= \sum_{i=1}^3\left|Z_{\alpha}^D\right|^2\left|I_{\mathrm{DP}_i}\right|^2\left(1 + \gamma\right),
\end{aligned}
\end{equation}
\end{widetext}
and finally Eq. (\ref{eq:P_t}) can be rewritten as
\begin{equation}
    P_T = \frac{\left(1+\gamma\right)}{2}\mathrm{Re}\left(Y_w^*\right)\left|Z_\alpha^D\right|^2\sum_{i=1}^3\left|I_{\mathrm{DP}_i}\right|^2= P_{0} \left(1 + \gamma\right)
\end{equation}
where $P_0$ is the power in the case in which both conditions $(i)$ and $(ii)$ are satisfied. Thus, $\gamma$ quantifies the error introduced by the deviations in the non-compliance with the ideal conditions for the method to be carried out. Since $\gamma$ depends directly on $\bar{I}_{\mathrm{DP}}$, the error is directly related to the direction considered for $\hat{n}$. 

%%%%%%%%%%%%%%%%%%%%%%%%%%%%%%%%%%%%%%%%%%%%%%%%%%%%%%%%%%%%%%%%%%%%%%%
\section{Electromagnetic design}\label{sec:emdesign}
%%%%%%%%%%%%%%%%%%%%%%%%%%%%%%%%%%%%%%%%%%%%%%%%%%%%%%%%%%%%%%%%%%%%%%%

This section details the proposed resonant cavities for the detection of DPs using three orthogonal degenerate modes. Three cavities with canonical geometries have been studied: cubic, spherical and cylindrical. Each of these incorporates three coaxial ports, and auxiliary tuning systems to slightly adjust the frequencies of the three modes for operating at the same frequency. The coaxial ports used in the simulations have the following features: outer radius $4.1$ mm, inner radius $1.27$ mm and a relative electric permittivity of $\varepsilon_r=2$, resulting in a characteristic impedance of $49.7 \, \Omega$. The cavities have been electromagnetically designed and simulated with CST Studio Suite \cite{CST}.

\subsection{Study of the cubic and spherical cavities}

\begin{figure}[h!]
    \centering
    \includegraphics[width=0.8\linewidth]{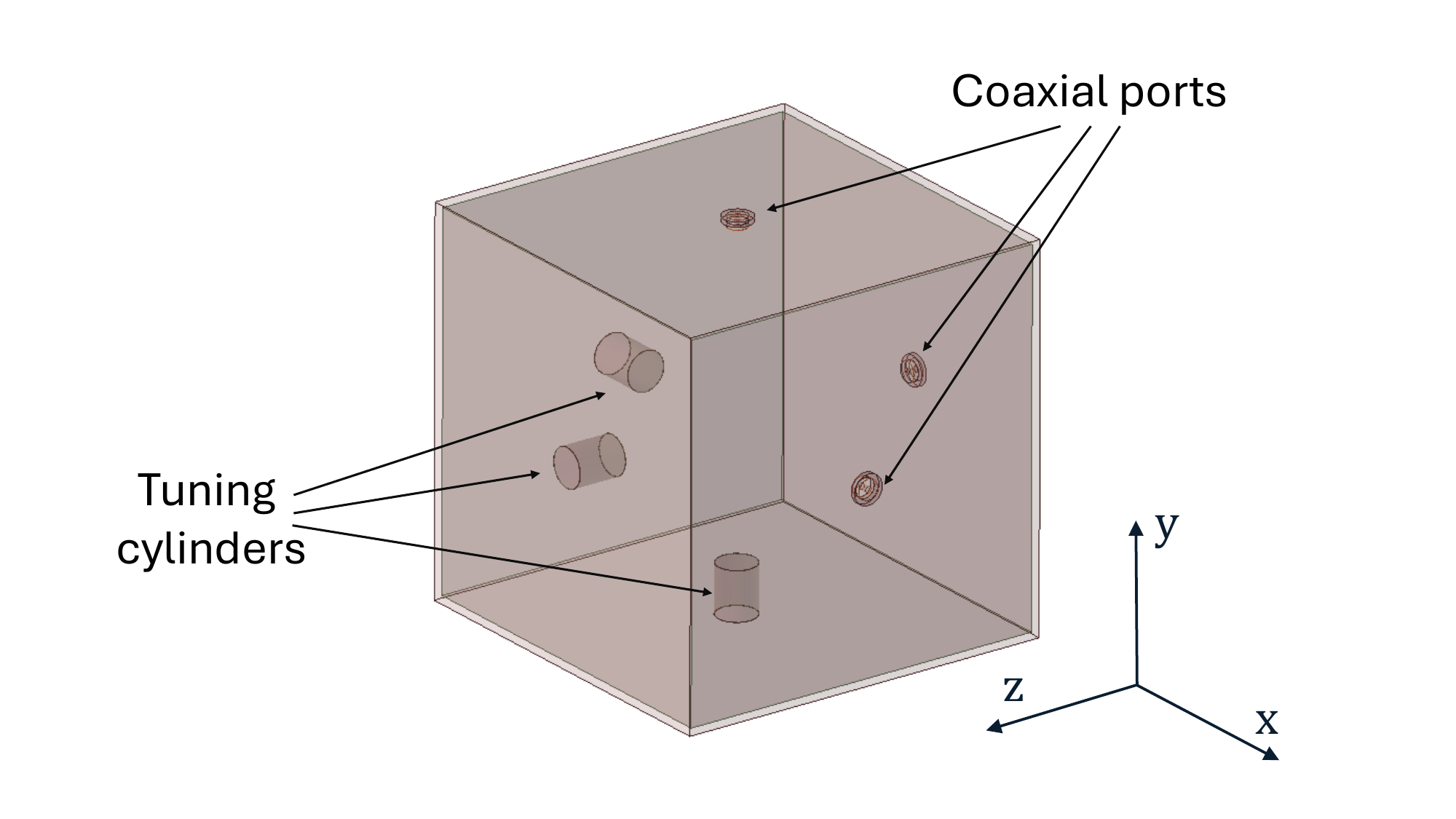}
    \caption{3D model of the cubic cavity for simulations with the coaxial ports and the tuning screws. The metallic tuning screws radius is 4 mm.}
    \label{fig:model_cubic}
\end{figure}

As a preliminary concept to assess the viability of this idea, the simplest case of a cubic cavity with a side length of $75$ mm has been designed, as depicted in Fig. \ref{fig:model_cubic}. This structure was previously explored for the detection of high-frequency gravitational waves \cite{Pablo_HFGW}.

This cavity incorporates three perpendicular coaxial ports, each oriented along the $x$-, $y$-, and $z$-axes in order to detect three degenerate modes, which in this case are $\textrm{TE}_{011}$, $\textrm{TE}_{101}$, and $\textrm{TM}_{110}$. The resonant frequency for these modes in this design is $f_r = 2.8259$ GHz. It is evident that in a realistic manufactured cavity the imperfections of the fabrication process will generate slight changes in the three resonant frequencies, which will have to be properly compensated with three tuning systems. This tuning mechanism consists of three cylindrical metallic screws positioned on the center of the opposite sides from the coaxial ports, which are properly inserted or extracted in order to modify independently the modal resonant frequencies. Hence, for the simulations made regarding the cubic geometry, the tuning screws have been slightly introduced into the cavity in order to study its impact in the final result, since some tolerances may appear in a realistic manufactured device as it has been mentioned previously.  Moreover, it is important to note that when modifying the port coupling, the resonant frequencies of the modes will be slightly modified, and thanks to the tuning system, the same frequency for the three modes can be re-achieved. As in any tuning system with moving parts, care must be taken in order to reduce the possible leakage and therefore avoiding a significative reduction of the cavity quality factor.\\

\begin{figure}[h!]
    \centering
    \includegraphics[width=0.9\linewidth]{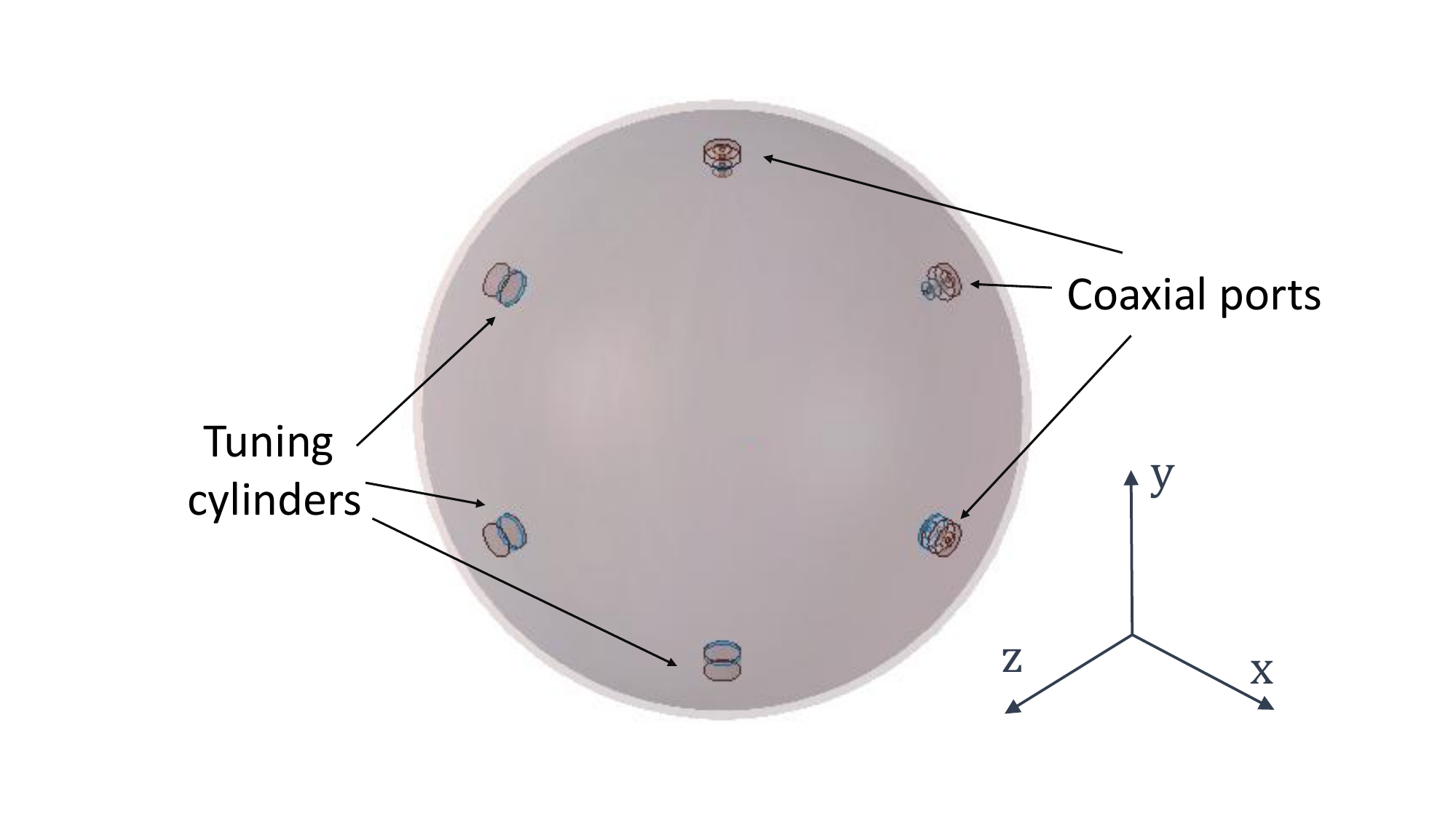}
    \caption{3D model of the spherical cavity for simulations with the coaxial ports and the tuning screws. The metallic tuning screws radius is 4 mm.}
    \label{fig:model_spherical}
\end{figure}

A similar study has also been carried out with a spherical cavity of radius $130.912$ mm, obtaining the $\mathrm{TM}_{011}$, $\mathrm{TM}_{111}$ (even), and $\mathrm{TM}_{111}$ (odd) modes at $f_r=2$~GHz. The 3D model of this cavity is shown in Fig. \ref{fig:model_spherical}. As in the case of the cubic cavity, the three ports are positioned orthogonally and the tuning metallic screws are located on the opposite surface to them.

\subsection{Study of the cylindrical cavity}
%%%%%%%%%%%%%%%%%%%%%%%%%%%%%%%%%%%%%%%%%%%%%%%%%%%%%%%%%%%

\begin{figure}[h!]
    \centering
    \includegraphics[width=0.9\linewidth]{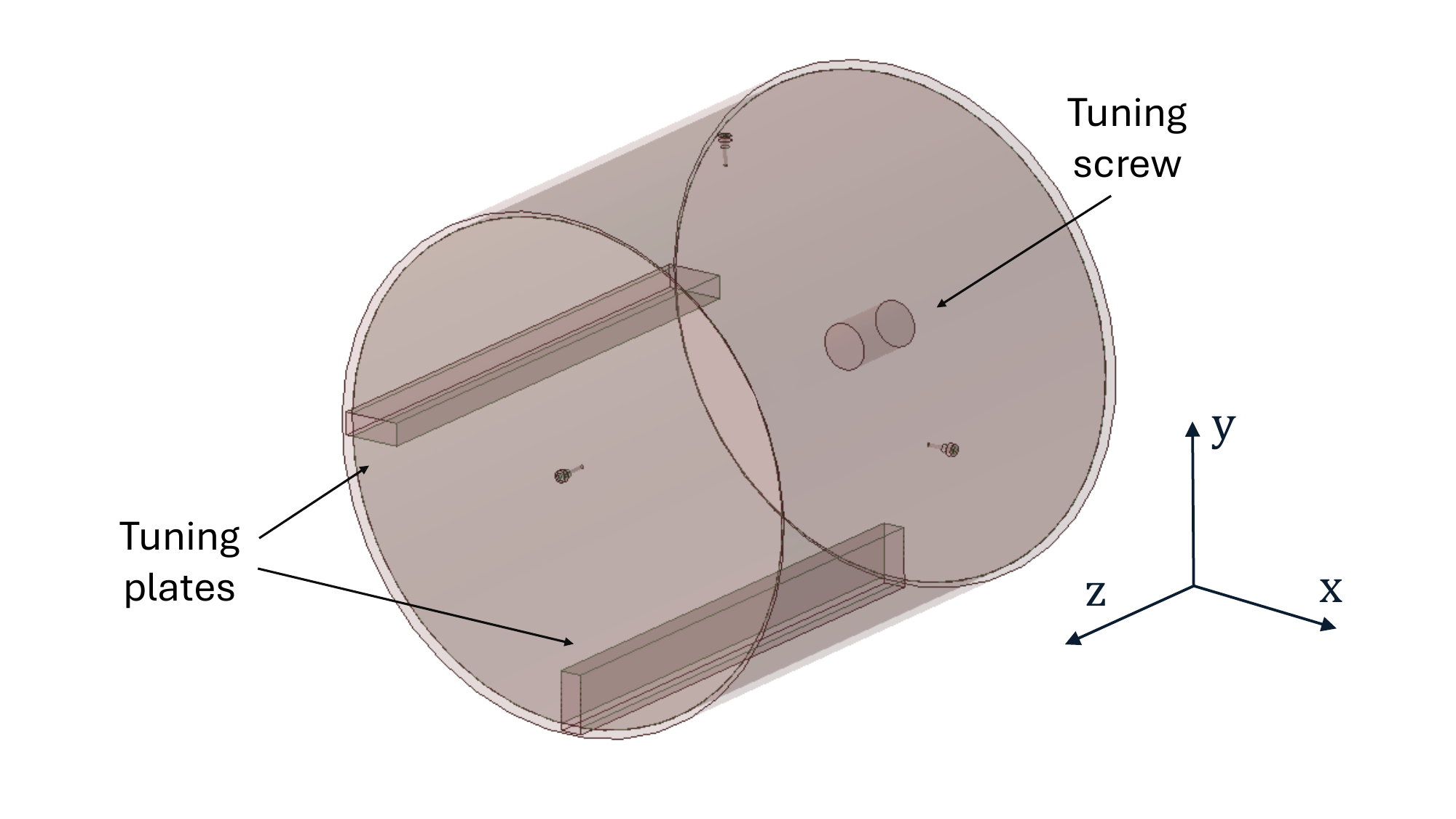}
    \caption{3D model of the cylindrical cavity with the coaxial ports, two metallic rectangular tuning plates and one metallic tuning screw. The tuning plates thickness and screw radius are 10 mm. The tuning plates length is the same as the cavity length.}
    \label{fig:model_cylindric}
\end{figure}

While cubic and spherical geometries are interesting examples which provide a more intuitive insight, typical experiments are designed to optimize the available volume of the magnet \footnote{No magnet is needed for DP detection. However, this method is intended for the simultaneous search for DPs and axions, where a magnet is needed.}. For such purpose, cylindrical cavities are more suitable for use in axion or DP detection experiments. The designed cylindrical cavity has a radius of $110$ mm and a length of $224.5$ mm. A schematic of this cavity is shown in\:Fig. \ref{fig:model_cylindric}. \\

The modes used in this cavity are the degenerate $\textrm{TE}_{111}$ with $x$- and $y$-polarization, respectively, and the $\textrm{TM}_{010}$ mode, whose electric field is oriented towards $z$-direction. The tuning system in this design consists, on one hand, of two identical metallic plates that are oriented $90^{\circ}$ from each other and whose length is equal to that of the cavity. These plates allow mainly to affect the $\textrm{TE}_{111-x}$ and $\textrm{TE}_{111-y}$ modes, whose resonant frequency variation is achieved by introducing or extracting the plates radially into the cavity. On the other hand, the frequency of the $\textrm{TM}_{010}$ mode is modified by means of a metallic screw located on the center of one base of the cylindrical resonator, which is inserted or extracted and affects the maximum of the modal electric field. The design of this cavity was made to have an operational frequency for the three degenerate modes at $f_r = 1.0414$ GHz, and it is worth to note that the dimensions of this cavity were not chosen to make the three modes to match directly in frequency. On the contrary, the $\mathrm{TM_{010}}$ is displaced in frequency with respect to the modes $\mathrm{TE}_{111-x}$ and $\mathrm{TE}_{111-y}$. This was made to observe the impact of having to introduce considerably the tuning elements to make the modes match in frequency, and in order to illustrate a more general example where no initial restriction is imposed over the cavity length. However, in order to compare with the case of having a cylindrical cavity whose three operational modes naturally coincide in frequency, another cylindrical cavity has also been designed whose radius is the same as the last cylindrical cavity but the length has been modified to match the frequencies of the $\textrm{TE}_{111-x}$, $\textrm{TE}_{111-y}$ and $\textrm{TM}_{010}$ modes. The dimensions of this cavity are therefore radius 110 mm and length $223.69$ mm.

An important remark here is that the fine-tuning system for obtaining the same resonant frequency for the three modes can also provide, at some extent, a scanning range for extending the search in a small frequency range. This has been studied for the three geometries, and a tuning frequency range over 1\% of the resonant frequency has been obtained. Future work is envisaged to increase thi scanning range.

\begin{figure}[t]
    \centering
    \includegraphics[width=0.99\linewidth]{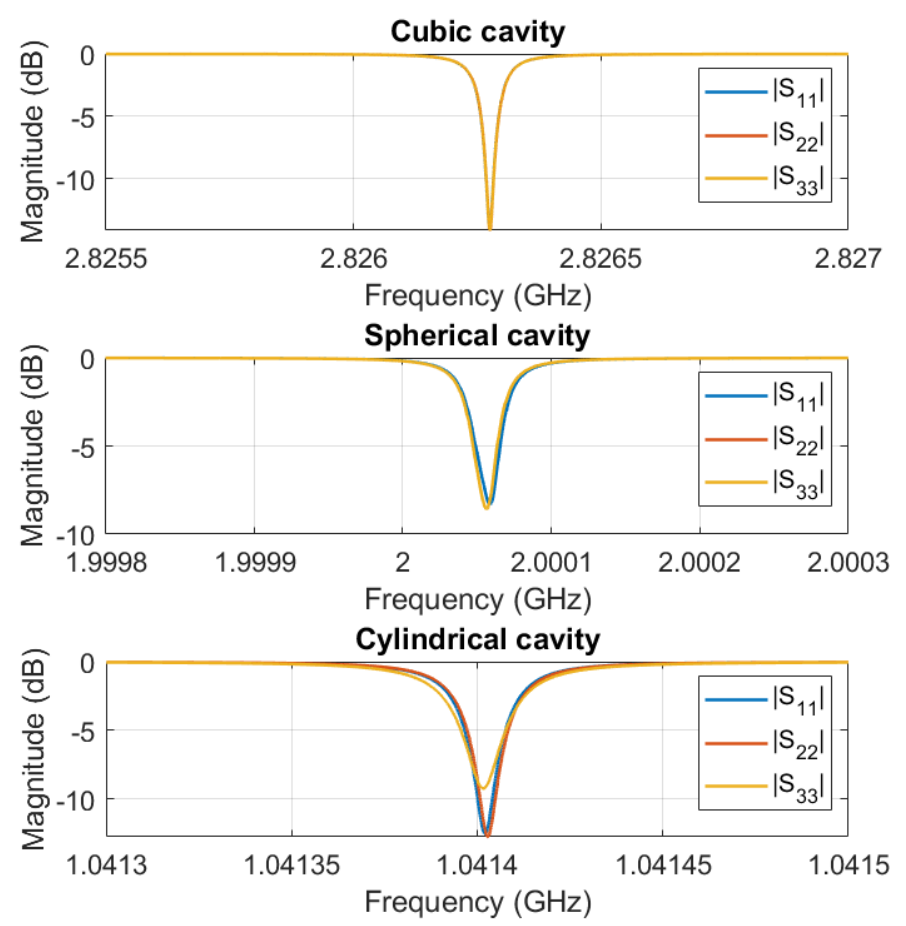}
    \caption{Reflection coefficients of the three modes for the cubic (upper plot), spherical (middle plot) and cylindrical (lower plot) cavities from frequency domain simulations, assuming a copper conductivity of $\sigma = 10^9$ S/m. The three ports in the cubic cavity have $\beta_1=\beta_2=\beta_3 = 0.67$, in the spherical cavity $\beta_1=\beta_2=\beta_3 = 0.46$ and in the cylindrical cavity $\beta_1=\beta_2=1.62$ and $\beta_3=2.06$.}
    \label{fig:Sparam}
\end{figure}

\section{Numerical Results}\label{sec:formfactorcalculation}

\subsection{Electrical response of the cavities}
Fig. \ref{fig:Sparam} shows the reflection coefficient at the operational frequencies for each port in the cubic, spherical and cylindrical cavities with tuning elements to illustrate that the same frequency can be achieved for the three modes. 

\subsection{Form factor calculations}
\label{sec:form_factor_calculations}
The computing of the DP form factor in the three studied cavities involves the assumption of different directions of the vector $\hat{n}$. Consequently, the individual form factors of each degenerate resonant mode are calculated as a function of the angle of $\hat{n}$ with respect to the coordinates axes ($\theta$ and $\varphi$, in spherical coordinates), and the total form factor can be obtained as the sum of the individual contributions of each mode (assuming that the detected power is coherently summed, which was demonstrated in section \ref{sec:birme}). \\

%\begin{figure} [ht]
  %  \centering
   % \includegraphics[width=0.85\linewidth]{images/nvector.pdf}
    %\caption{3D representation of the possible directions that $\hat{n}$ can take. The vector $\hat{n}$ is the conventional unit radial vector in spherical coordinates. The blue points represent their projection on a spherical surface of radius equal to one. Two black arrows representing two different $\hat{n}$ are shown as an example.}
    %\label{fig:DP_n}
%\end{figure}

\begin{figure*}[t]

    \centering
    \renewcommand{\arraystretch}{1.8} % Aumenta el espacio vertical en un 50%
    \begin{tabular}{| >{\centering\arraybackslash}m{0.47\textwidth} | >{\centering\arraybackslash}m{0.47\textwidth} |}
        \hline
        \multicolumn{1}{|c|}{\Large $\textrm{$C_1$ ($\textrm{TE}_{111-x}$)}$} & \multicolumn{1}{c|}{\Large $\textrm{$C_2$ ($\textrm{TE}_{111-y}$)}$} \\
        \hline
        \parbox[c]{\linewidth}{\centering \resizebox{\linewidth}{!}{\includegraphics{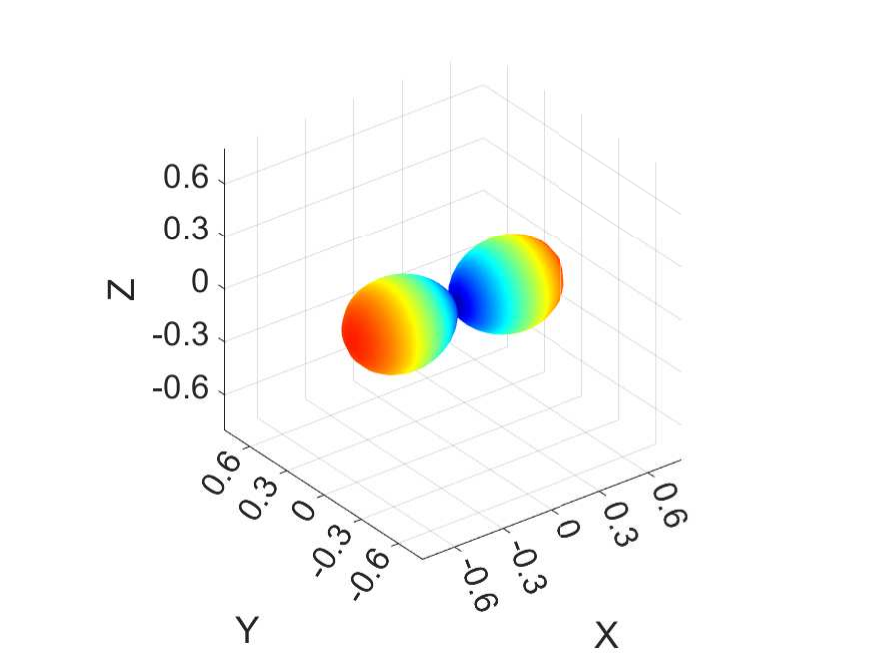}}\vspace{0.2cm}}  & 
        \parbox[c]{\linewidth}{\centering \resizebox{\linewidth}{!}{\includegraphics{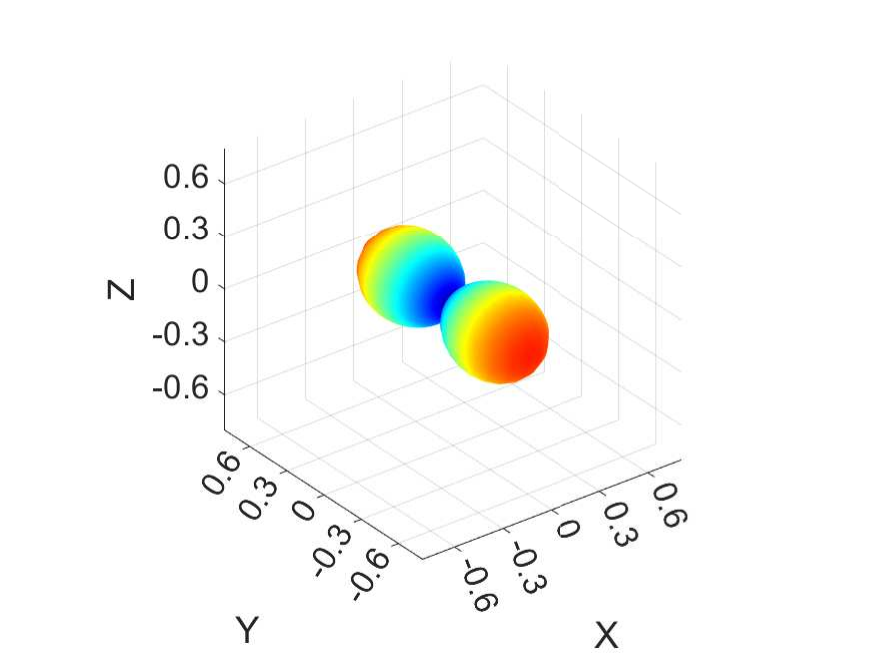}}\vspace{0.2cm}} \\
        \hline
        \multicolumn{1}{|c|}{\Large  $\textrm{$C_3$ ($\textrm{TM}_{010}$)}$} & \multicolumn{1}{c|}{\Large $\textrm{$C_T$}$} \\
        \hline
        \parbox[c]{\linewidth}{\centering \resizebox{\linewidth}{!}{\includegraphics{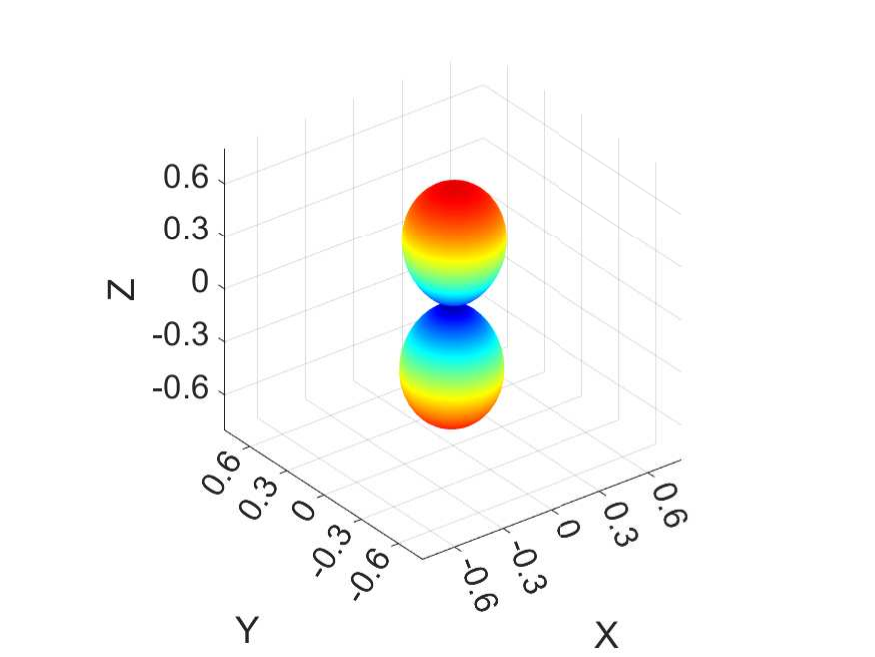}}\vspace{0.2cm}} & 
        \parbox[c]{\linewidth}{\centering \resizebox{\linewidth}{!}{\includegraphics{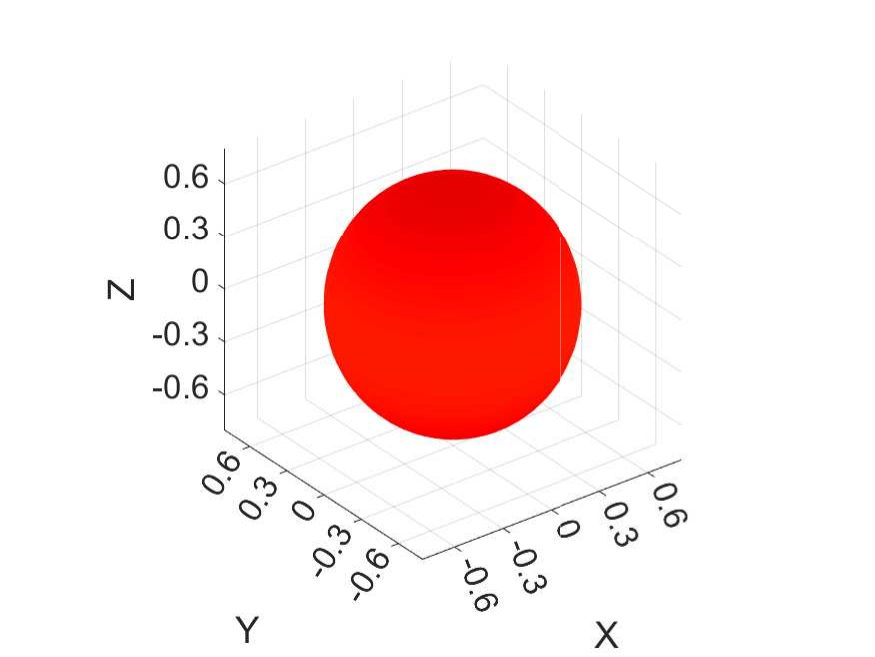}}\vspace{0.2cm}} \\
        \hline
    \end{tabular}

    % Agregar barra de color global debajo de la tabla
    \vspace{0.5cm}
    \includegraphics[width=0.5\textwidth]{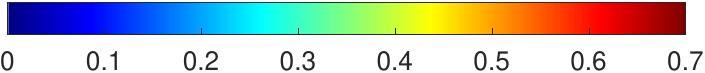} % Colorbar separado

    \caption{Form factor calculations for each mode ($C_1, C_2$ and $C_3$) and the total ($C_T$) in the cylindrical cavity depending on the direction of $\hat{n}$. Both the radius and the color scale of each plot indicate the value of the form factor. The same form factor maps were obtained for the cubic and spherical geometries, considering the modes $\mathrm{TE}_{011}, \mathrm{TE}_{101},  \mathrm{TM}_{110}$ for the cubic cavity and the modes $\mathrm{TM_{011}}, \mathrm{TM_{111}}\, \mathrm{(even)}, \mathrm{TM_{111}}\, \mathrm{(odd)}$ for the spherical one.}
    \label{tab:cylindrical_cavity}
\end{figure*}

To verify the experiment sensitivity in all directions, the form factor for a given number of the possible orientations of $\hat{n}$ contained within a sphere, and for each resonant mode has been obtained. The results observed in Fig. \ref{tab:cylindrical_cavity} represent the form factor map for each mode ($C_1$, $C_2$ and $C_3$) as well as the total form factor $C_T = C_1 + C_2 +C_3$. In order to avoid redundancy, results shown in Fig. \ref{tab:cylindrical_cavity} are the ones obtained for the cylindrical cavity with tuning elements introduced, but the exact same form factor maps were obtained for the cubic and spherical geometries.\\
%provided that the $Q_L$ is approximately the same for the three modes \footnote{For the cubic and spherical cavities, it is trivial to see that this condition is fulfilled. For the cylindrical geometry there is a slightly difference in the unloaded quality factors, being $1.5\cdot10^5$ for the $\mathrm{TM_{010}}$ and $1.56\cdot10^5$ for both $\mathrm{TE_{111}}$.}

From the plots, it can be observed that, as expected, the more parallel $\hat{n}$ to the electric field components of a given mode, the higher the form factor due to the greater contribution of such mode. For example, for the $\textrm{TE}_{111-x}$ mode, whose electric field has mainly $x$-component, the maximum value of $C_{\mathrm{DP}}$ is obtained when $\hat{n}$ is oriented to the $x$-axis ($\varphi = 0, \pi$ and $\theta = \pi/2$), and likewise for the $\textrm{TE}_{111-y}$ mode when $\hat{n}$ is oriented to the $y$-axis ($\varphi = \pi/2, 3\pi/2$ and $\theta = \pi/2$). Finally, in the case of $\textrm{TM}_{010}$ mode, when $\hat{n}$ is as parallel as possible to the $z$-direction ($\theta = 0, \pi$), the value of $C_3$ is maximum, decreasing when $\theta$ varies. For a fixed value of $\theta$, the same value is obtained for any angle $\varphi$. \\

The next step is to analyze the perturbation generated by the irregularities derived from the tuning systems. For such purpose, we have defined the relative irregularity $\Lambda$ as the difference between the maximum and the minimum values of $C_{T}$ in percentage, that is,
\begin{equation}
    \Lambda(\%)=\frac{\mathrm{max}(C_T)-\mathrm{min}(C_T)}{\mathrm{max}(C_T)} \, 100.
\end{equation}

The source of this irregularity comes from the effect of the tuning systems, which perturb the electric field pattern differently depending on their geometric structure and the depth at which they are inserted into the cavity. The results for the computation of the maximum form factor $C_T$ and irregularity $\Lambda$ are shown in Table \ref{tab:C_T_results}. As it can be seen, the best performance is observed in the spherical cavity, due to the symmetry between its three orthogonal modes. For the case of the cubic and cylindrical cavities, the results in the form factor are very similar. Since the perturbations introduced by the plates in the $x-$ and $y-$axes of the cylindrical geometry are much more relevant than the ones introduced by the screw in the $z-$direction, the irregularity is higher for this geometry. Moreover, this disturbance might introduce an uncertainty in the detected power, as it will be discussed in section \ref{sec:uncertainty}. As a consequence, taking into account Eq. (\ref{eq:chi}), if one assumes the first approach for the DP polarization (DP permanent polarization, $C_\mathrm{DP} = 0.0025C_a$) the improvement in sensitivity (in the cubic and spherical cavities) for a given integration time is a factor 20, whilst if one assumes a given sensitivity, the improvement in integration time is a factor $1.6\,\cdot\,10^5$. On the other hand, if one assumes the second approach for the DP polarization (in which $\hat{n}$ is random and $C_\mathrm{DP}=(1/3) C_a$), the improvement in sensitivity for a given integration time is a factor of 1.7, whilst the improvement in integration time for a given sensitivity is a factor 9. As a result, the proposed setup implies robust improvements for both approaches of the DP polarization, but in any case, the error introduced by the tuning systems must be taken into account. This error is analyzed for the three geometries in the next subsection \ref{sec:uncertainty}.

\begin{table}[h]
\centering
\begin{tabular}{ c | c | c }
\hline
\textbf{Geometry} & max($C_T$) & $\Lambda$ ($\%$)\\
\hline
Cubic & 0.67 & 2.5 \\ 
\hline
Spherical & 0.72 & 0.002 \\
\hline
Cylindrical & 0.69 & 1.24 \\
\hline
Cylindrical with tuning & 0.67 & 5.6 \\

\hline
\end{tabular}
\caption{Maximum values of $C_T$ and values of $\Lambda$ obtained for the four studied cavities.}
\label{tab:C_T_results}
\end{table}

\begin{figure*}[t]

    \centering
    \renewcommand{\arraystretch}{1.8} % Aumenta el espacio vertical en un 50%
    \begin{tabular}{| >{\centering\arraybackslash}m{0.4\textwidth} | >{\centering\arraybackslash}m{0.4\textwidth} |}
        \hline
        \multicolumn{1}{|c|}{\Large $\gamma$ in the cubic cavity} & \multicolumn{1}{c|}{\Large $\gamma$ in the cylindrical  cavity with tuning} \\
        \hline
        \parbox[c]{\linewidth}{\centering \resizebox{\linewidth}{!}{\includegraphics{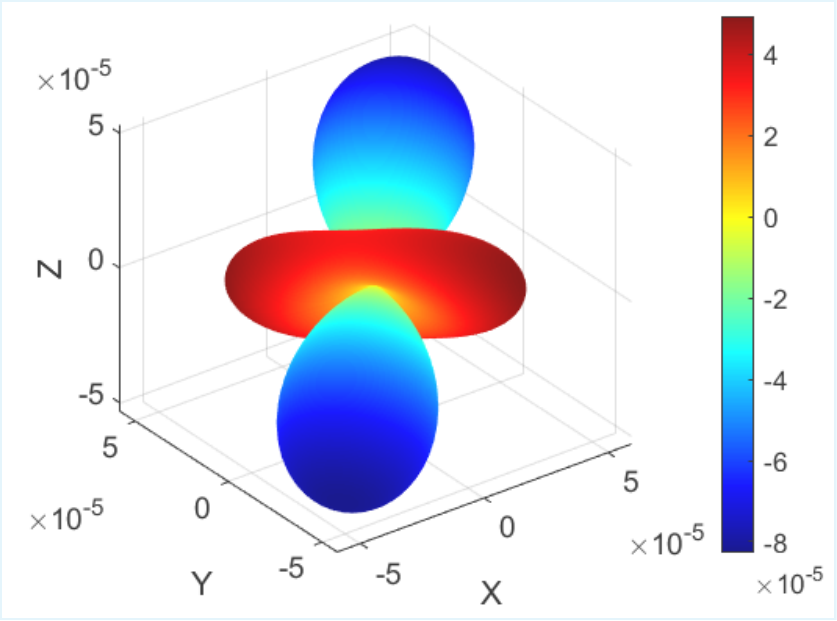}}\vspace{0.2cm}}  & 
        \parbox[c]{\linewidth}{\centering \resizebox{\linewidth}{!}{\includegraphics{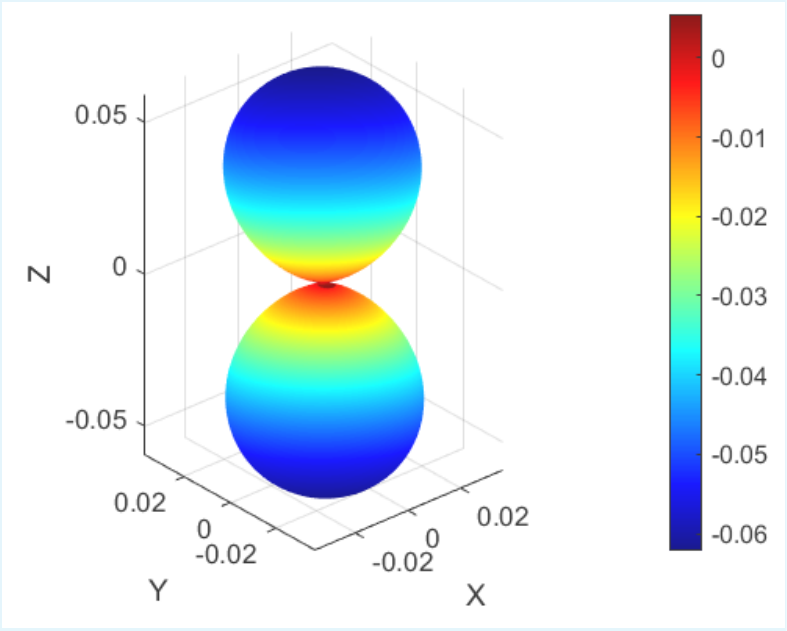}}\vspace{0.2cm}} \\
        \hline
    \end{tabular}

    \caption{Error $\gamma$  for cubic and cylindrical with tuning cavities as a function of the polarization vector $\hat{n}$. Both the radius and the color scale of each plot indicate the value of $\gamma$. The order of magnitude obtained for the spherical geometry is the same as for the cubic one, since both conditions of the admittance matrix are also fulfilled in the spherical case.}
    \label{tab:gamma_error}
\end{figure*}

In addition, this setup can also allow to reconstruct the original direction of the polarization vector $\hat{n}$ during the measurement, always provided that the DP has been detected and the variation period of the DP polarization is much larger than the integration time, making this directional study only valid for the first approach of the DP polarization. One can find that the $\left(\theta, \varphi\right)$ angles of the incoming DP signal can be calculated from the form factors or from the detected power in each port as \\

\begin{equation}
    \begin{aligned}
        \theta &= \arccos\left(\sqrt{\frac{P_{w_3}}{P_{T}}}\right); \\
        \varphi &= \arctan\left(\sqrt{\frac{P_{w_2}}{P_{w_1}}}\right).
    \end{aligned}
\end{equation}
To accomplish this, signals at each port must be electronically processed individually. Given that the DP frequency is already known in this approach, it is possible to integrate the signal more time, thereby enhancing the SNR. Additionally, the distinction between an axion and a DP can be determined by disabling the magnet and observing if the signal persists, indicating the possibility of the DP detection.
\renewcommand{\arraystretch}{1.5}
\begin{table*}
  \centering
  \begin{tabular}{ c | c | c}
    \hline
    \textbf{Geometry}   & $\chi$ improvement with respect to $\cos^2\left(\theta\right)=0.0025$              & $\chi$ improvement with respect to  $\langle\cos^2\left(\theta\right)\rangle=1/3$         \\
    \hline
    Cubic      & $20.0\pm 0.4$   & $1.73\pm0.03$ \\
    \hline
    Spherical      & $20.000 \pm 0.002$   & $1.7321\pm 0.0002$ \\
    \hline
    Cylindrical& $ 20.0 \pm 0.2 $             & $1.72 \pm 0.02$      \\
    \hline
    Cylindrical with tuning& $20\pm 1$             & $1.7\pm 0.1$     \\
    \hline
  \end{tabular}
  \caption{Improvement in sensitivity $\chi$ while maintaining the integration time $\Delta t$ constant, depending on the value of $\cos^2\left(\theta\right)$ and on the geometry considered, with the total error $\epsilon$.}
  \label{tab:sensitivity_results}
\end{table*}
\renewcommand{\arraystretch}{1.5}
\begin{table*}
  \centering
  \begin{tabular}{ c | c | c}
    \hline
    \textbf{Geometry}   & $\Delta t$ improvement with respect to $\cos^2\left(\theta\right)=0.0025$              & $\Delta t$ improvement with respect to  $\langle\cos^2\left(\theta\right)\rangle=1/3$         \\
    \hline
    Cubic       & $\left(1.60\pm 0.01\right)\cdot 10^5$   & $9.00\pm 0.07$ \\
    \hline
    Spherical      & $\left(1.6000\pm0.0004\right)\cdot 10^5$   & $9.000\pm0.003$ \\
    \hline
    Cylindrical & $\left(1.60\pm0.06\right)\cdot 10^5$            & $9.0\pm0.3$    \\
    \hline
    Cylindrical with tuning & $\left(1.6\pm0.4\right)\cdot 10^5$             & $9\pm 2$     \\
    \hline
  \end{tabular}
  \caption{Improvement in integration time $\Delta t$ while maintaining the sensitivity $\chi$ constant, depending on the value of $\cos^2\left(\theta\right)$ and on the geometry considered, with the total error $\epsilon$.}
  \label{tab:integration_time_results}
\end{table*}
\subsection{Numerical analysis of the uncertainty caused by the error} \label{sec:uncertainty}
In this subsection we calculate the uncertainty introduced in the application of the technique proposed in this article. For the cases in which the tuning elements are not considerably introduced into the cavities, it is intuitive to see that the method is applicable. However, for the cylindrical cavity with tuning elements, the tuning plates are inserted 1 mm, and the tuning screw must be inserted 7.95 mm into the resonator in order to bring the $\mathrm{TM_{010}}$ mode frequency into coincidence with the one of the $\mathrm{TE_{111}}$ modes. This asymmetric perturbation alters the cavity’s admittance matrix and thus impacts the coherent summation of the extracted signals.\\

In Fig. \ref{tab:gamma_error} the error $\gamma$ is shown for the cubic and cylindrical (with tuning elements) geometries having into account the different directions of $\hat{n}$ (the plot for the spherical cavity and for the cylindrical geometry without tuning elements has been omitted since the results are similar to the ones for the cubic cavity). While for the cubic, spherical and cylindrical without tuning cavities the error is four orders of magnitude smaller than the maximum form factor achieved, for the cylindrical cavity with tuning elements it is one order of magnitude smaller. However, even with tuning elements the relative error of the final form factor is still very small, implying that this method has still a lot of margin to consider a major impact of the tuning elements in the final geometry. \\

In this way, $\gamma$ can be considered an uncertainty over the final form factor after the summation process. This translates into an uncertainty in the sensitivity and integration time improvement estimation made previously (discussed in subsection \ref{sec:form_factor_calculations}) for which the error due to the form factor irregularity ($\Lambda$) has also to be taken into account. For this purpose, the total error is defined as $\epsilon = \sqrt{\gamma^2+\Lambda^2}$, and this parameter affects the improvement estimation values, which are shown in Tables \ref{tab:sensitivity_results} and \ref{tab:integration_time_results}. By looking at these results, it can be concluded that the method discussed here works for the three geometries considered, even when introducing the tuning systems in the cylindrical geometry. The improvements in both $\chi$ and $\Delta t$ are very robust with respect to any of the two current approaches for the DP form factor, implying that the method proposed in this work for detecting the DP is the most competitive one. In addition, as mentioned previously, there is still a lot of margin to introduce more the tuning elements, in order to be able to increase the frequency scan.\\

Finally, as Eq. (\ref{eq:chi}) shows, not only the form factor, but the volume of the cavity has a high impact in the sensitivity. The analyzed cavities present different volumes, and therefore, in order to ensure a fair comparison, a realistic case with a bore radius of 110 mm has been assumed. Three cavities of the geometries discussed in the article that fit within that bore have been compared. These are a cubic cavity of 155.56 mm side, a spherical cavity of 110 mm in radius and finally the cylindrical cavity presented in section \ref{sec:emdesign}, whose dimensions (radius $110$ mm and length $223.69$ mm) naturally match the frequency of the degenerate modes. Results can be seen in Table \ref{tab:comparison_VC}. \\

\begin{table}[h]
\centering
\begin{tabular}{ c | c | c }
\hline
\textbf{Geometry} & $V\: \mathrm{(\cdot 10^{-3}\, m^3)}$ & $VC_T \:\mathrm{(\cdot 10^{-3}\, m^3)}$\\
\hline
Cubic & $3.781$ & $2.533$ \\ 
\hline
Spherical & $5.604$ & $4.035$ \\
\hline
Cylindrical & $8.533$ & $5.888$ \\
\hline
\end{tabular}
\caption{Volume and $VC_T$ factor of three cavities whose maximum dimension is fixed for a bore of radius of 110 mm.}
\label{tab:comparison_VC}
\end{table}

As expected, the geometry that takes the most profit from the bore space is the cylindrical one. The method studied in this work properly applies to the cylindrical geometry, and therefore, this case is the most interesting one, since the majority of axion experiments are using cylindrical cavities. Moreover, the volume of a cylindrical cavity can be increased in length, but in the other side, this will produce a separation of the $\mathrm{TE}_{111}$ modes and the $\mathrm{TM}_{010}$ mode, thus leading to a more invasive tuning in order to reach the degenerate state and, therefore, increasing the error.

%%%%%%%%%%%%%%%%%%%%%%%%%%%%%%%%%%%%%%%%%%%%%%%%%%%%%%%%%%%%%%%%%%%%%%%%%%%%%%%
\section{Conclusions}
%%%%%%%%%%%%%%%%%%%%%%%%%%%%%%%%%%%%%%%%%%%%%%%%%%%%%%%%%%%%%%%%%%%%%%%%%%%%%%%

This work explores the idea of using degenerate and orthogonal EM modes in haloscopes to improve the detectability of DPs. Since the DP polarization is unknown, we have demonstrated that the use of three degenerate modes whose electric field profile is oriented in three orthogonal directions of space considerably improves the form factor of a DP experiment until the point of obtaining the axion form factor. On one hand, four cavities, one cubic, one spherical and two cylindrical have been designed and simulated, three of them incorporating precise tuning systems to match the frequency of the three chosen orthogonal modes, and one of the cylindrical cavities with dimensions to match naturally the frequency of the three modes. From the electric field profile, a mapping of the DP form factor for each mode and the total DP form factor has been carried out, verifying that, by coherently summing the signals of the three ports, the total DP form factor is practically the same as that of the axion. The isotropy of the system for the DP form factor has been calculated, obtaining robust results: for the spherical cavity, the total form factor map irregularity is of $0.002\%$, while for the cubic and cylindrical geometries the irregularities obtained are of $2.5\%$, $1.24\%$ and $5.6\%$, respectively, implying a quasi isotropic response to the DP polarization. In addition, this method opens the possibility to determine the direction of the DP polarization, for which it is necessary to monitor each port separately and to measure the individual power obtained by each of them. After all, a remarkable improvement in sensitivity as well as in integration time is obtained, depending on the approach considered for the DP polarization: when assuming that $\hat{n}$ is fixed, the sensitivity of the experiment improves by a factor 20; whilst when averaging over all possible directions of $\hat{n}$, the sensitivity improves by a factor $1.7$. The same reasoning can be followed for calculating the integration time improvement for reaching a certain sensitivity, obtaining a factor $1.6 \cdot 10^5$ of improvement when $\hat{n}$ is assumed to be constant, whilst when averaging over all possible directions of $\hat{n}$ a factor 9 is found. We have also developed the conditions necessary to be able to coherently sum the signals of the modes for the three designs, finding that by introducing tuning systems to match the frequencies, the sensitivity and integration time improvement factors mentioned previously have an associated uncertainty. Because of this, the example of the cylindrical cavity with tuning elements has been of great importance, as it was purposely designed so that mode $\mathrm{TM}_{010}$ does not match in frequency with modes $\mathrm{TE}_{111,x}$ and $\mathrm{TE_{111,y}}$ in order to observe the impact of the tuning elements. As a result, it has been found that the method also works properly, finding a low relative uncertainty of the form factor. This implies that there is still a lot of margin for introducing the tuning elements as in other DP and axion experiments, in order to increase the frequency scan range. Moreover the proposed setups allow for a simultaneous detection of both axions and DPs, implying that a experiment which is originally built up for detecting axions is able, if the tuning system is designed for making three orthogonal modes coincide in frequency, to detect the DP with maximum form factor. In this way, we would like to remark that, since the vast majority of axion haloscopes employ cylindrical geometries, the technique developed in this work can be translated to any axion experiment with the mentioned geometry in order to achieve optimal sensitivity to DPs. Nowadays, to the knowledge of the authors, the method proposed in this work is the most effective one to search for DPs with a microwave resonant cavity.\\

\section{\label{acknowledgments} Acknowledgments}
This work was performed within the RADES group; we thank our colleagues for their support. Thanks also to Juan Monzó Cabrera for his valuable comments and discussions. The research leading to these results has received funding from the Spanish Ministry of Science and Innovation with the projects PID2022-137268NBC53 and PID2022-137268NA-C55, funded by MICIU/AEI/10.13039/501100011033/ and by “ERDF/EU”. J. Reina-Valero has the support of "Plan de Recuperación, Transformación y Resiliencia (PRTR) 2022 (ASFAE/2022/013)", founded by Conselleria d'Innovació, Universitats, Ciència i Societat Digital from Generalitat Valenciana, and NextGenerationEU from European Union.

\bibliography{references}

\end{document}